\documentclass{article}
\usepackage{amsmath}
\usepackage{amsfonts}
\usepackage{epsfig}

\usepackage[preprint]{spconfa4}
\usepackage{algorithmicx}
\usepackage{pifont}
\usepackage{algpseudocode}
\usepackage{algorithm}
\usepackage{lipsum}

\copyrightnotice{\textbf{978-1-7281-1331-9/20/\$31.00~\copyright 2020 IEEE}}

\let\OLDthebibliography\thebibliography
\renewcommand\thebibliography[1]{
  \OLDthebibliography{#1}
  \setlength{\parskip}{0pt}
  \setlength{\itemsep}{0pt plus 0.3ex}
}

\begin{document}\sloppy

\def\x{{\mathbf x}}
\def\L{{\cal L}}

\title{Voice-Indistinguishability: Protecting Voiceprint in Privacy-Preserving Speech Data Release}
%
\name{Yaowei Han$^{\ast}$, Sheng Li$^{\dagger}$, Yang Cao$^{\ddagger}$, Qiang Ma$^{\ddagger}$ and Masatoshi Yoshikawa$^{\ddagger}$}
\address{$^{\ast}${$^{\ddagger}$}Department of Social Informatics, Kyoto University, Kyoto, Japan\\$^{\dagger}$National Institute of Information and Communications Technology, Kyoto, Japan\\\hspace*{-4pt}E-mail: $^{\ast}$yaowei@db.soc.i.kyoto-u.ac.jp,$^{\dagger}$sheng.li@nict.go.jp,$^{\ddagger}$\{yang,qiang,yoshikawa\}@i.kyoto-u.ac.jp}

\maketitle
\begin{abstract}
\vspace{1pt}
With the development of smart devices, such as the Amazon Echo and Apple's HomePod, speech data have become a new dimension of big data.
However, privacy and security concerns may hinder the collection and sharing of real-world speech data, which contain the speaker's identifiable information, i.e., voiceprint, which is considered a type of biometric identifier.
Current studies on voiceprint privacy protection do not provide either a meaningful privacy-utility trade-off or a formal and rigorous definition of privacy.
In this study, we design a novel and rigorous privacy metric for voiceprint privacy, which is referred to as voice-indistinguishability, by extending differential privacy.
We also propose mechanisms and frameworks for privacy-preserving speech data release satisfying voice-indistinguishability.
Experiments on public datasets verify the effectiveness and efficiency of the proposed methods.
\end{abstract}

\section{Introduction}

With the advance of voice-based human-computer interaction and the development of smart devices, such as the Amazon Echo and Apple's Homepod, speech data have become a new dimension of big data.
The collection and sharing of real-world speech data not only enables the improvement of innovative services and products, such as Apple's Siri and Google Assistant but also fosters studies on intelligent algorithms.
\renewcommand{\thefootnote}{}
\footnotetext{The first three authors contributed equally to this work.}

However, privacy and security concerns may hinder the collection and sharing of real-world speech data.
First, speech data contain the speaker's identifiable information represented as \textit{voiceprint} (as analogous to fingerprints), which is considered a type of biometric identifier \cite{boles_voice_2017}.
Therefore, with the advent of the GDPR and increasing privacy concerns, the sharing of speech data is faced with significant challenges \cite{ nautsch_gdpr_2019}.
Second, exposing an individual's voiceprint may cause security risks.
Because voiceprints are used in many authentication systems \cite{boles_voice_2017, wechat2015}, an attacker may commit \textit{spooﬁng attacks} \cite{wu_spoofing_2015} to the voice authentication systems.
Additionally, a victim may suffer from \textit{reputation attacks}, as shown in a recent study on fake Obama speeches \cite{suwajanakorn2017synthesizing}.

Several methods \cite{justin2015speaker, qian2018hidebehind,srivastava2019evaluating, fang_speaker_2019} for anonymizing speakers' identities have been proposed to address these problems.
A naive way to protect voiceprint could be totally removing the voiceprint by converting the speech records into texts and releasing machine-generated voice by text-to-speech (TTS) software \cite{justin2015speaker}.
Although these methods can effectively protect voiceprint, the released speech data are not significantly useful because the \textit{speech data} not only include the linguistic information but also imply the speakers' characteristics, such as pitch, speaking rate, and emotion.
Some other methods \cite{qian2018hidebehind, srivastava2019evaluating} provide a better utility of the released speech data by utilizing voice conversion but the privacy protection is based on the hiding of the parameters used in the voice conversion.
If an attacker obtains these parameters, he can reverse engineer to the original voice.
A recent study \cite{fang_speaker_2019} proposed a speaker anonymization system by extending the k-anonymity model \cite{sweeney_k-anonymity:_2002}.
However, numerous studies \cite{narayanan_robust_2008, merener_theoretical_2012} on data privacy indicate that k-anonymity is less rigorous in many real-world applications because it is based on the assumptions of the attackers' background knowledge.

\textbf{Contributions.} 
In this study, for the first time, we design a novel and rigorous privacy metric for voiceprint privacy by extending differential privacy \cite{dwork2006calibrating} for privacy-preserving speech data release. 
First, we design a formal privacy notion for voiceprint, called \textit{voice-indistinguishability}, whose privacy guarantee does not depend on adversarial knowledge.
We use the state-of-the-art representation of voiceprint, i.e., the x-vector \cite{Snyder2018x}. 
The definition provides different degrees of indistinguishability for every two x-vectors according to their distance (i.e., similarity).
Second, we design a mechanism satisfying voice-indistinguishability and analyze the privacy guarantee provided by our definition, which is limiting the information an adversary can obtain.
Third, we propose privacy-preserving speech data release frameworks based on voice synthesis techniques.
We verify the effectiveness and efficiency of the proposed methods on public speech datasets.

\begin{figure}[t]
\begin{center}
\includegraphics[width=0.45\textwidth]{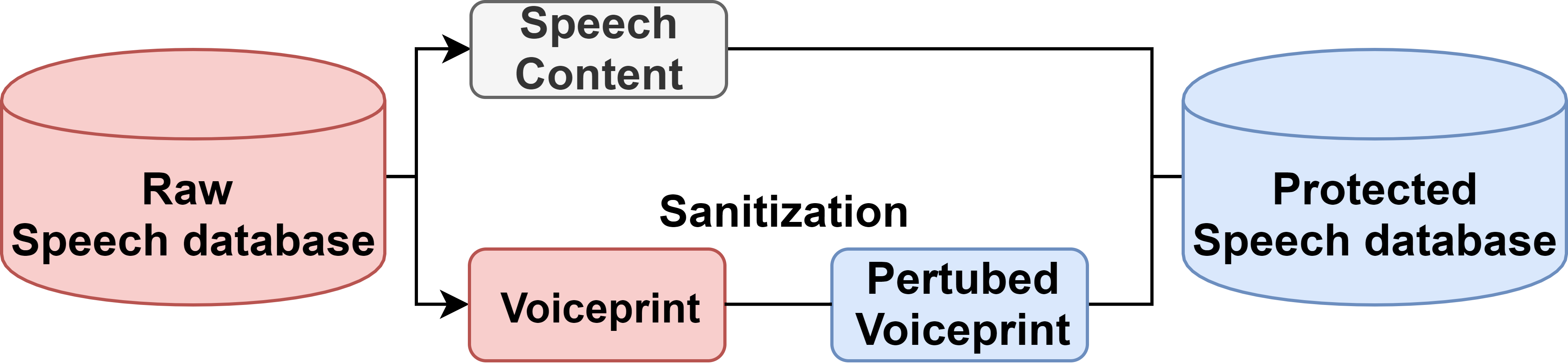}
\vspace{-5pt}
  \caption{Privacy-preserving speech data release (the red and blue indicate the sensitive and protected data, respectively).}
  \label{pro}
  \vspace{-15pt}
 \end{center}
\end{figure}

\section{Preliminaries}
\subsection{Voiceprint: what we intend to protect}
 
Voice is a hybrid of physiological and behavioral identifiers. As mentioned previously, voiceprint is an important biometric characteristic that only reflects the physiological identifier of voice. Voiceprint has the following characteristics: (1) Stableness: Voiceprint is relatively stable regardless of speakers' environmental and psychological changes; (2) Naturalness: Because voiceprint is a natural characteristic of voice, it is not limited to specific scenarios.

\subsection{Problem Setting}

In this study, we aim to protect voiceprint in the privacy-preserving speech data release.
In other words, the published speech database should ensure that the voiceprint embedded therein is indistinguishable. 
Figure \ref{pro} shows the privacy-preserving speech data release process. For each raw utterance, we extract its content and voiceprint, after which we sanitize the extracted voiceprint.
We use $\mathbb{D}={\{u_1, u_2, ..., u_n\}}$ to denote the speech database where $u_i$ represents the $i$th utterance of the speech database and $\mathcal{D} = {\{x_1, x_2, ..., x_n\}}, x \in \mathcal{X}$ to denote the voiceprint database extracted from $\mathbb{D}$, where $\mathcal{X}$ is the voiceprint domain. Note that $\mathbb{D}$ and $\mathcal{D}$ share the same size, $n$.

\section{Voice-indistinguishability}
\subsection{Privacy Definition and Mechanism}

When considering the achievement of the indistinguishability of voiceprint, differential privacy \cite{dwork2006calibrating} seems to be a satisfying solution.
Differential privacy is a state-of-the-art paradigm facilitating the secure analysis of sensitive data, owing to its strong assumption of an adversary's background knowledge and its smart setting of the privacy budget. 
However, differential privacy proposed for databases cannot be directly applied.
Alternatively, we consider a generalized differential privacy known as metric privacy \cite{chatzikokolakis2013broadening}. 
Metric privacy defines a distance metric between secrets and guarantees the indistinguishability of every two secrets to the extent that is proportional to the distance between each.

\textbf{Metric privacy:} A mechanism $K:\mathcal{X \to P(Z)}$ that satisfies $d_\mathcal{X}$-privacy, only if $\forall x, x' \in \mathcal{X}$
\begin{equation}
\setlength{\abovedisplayskip}{3pt}
\setlength{\belowdisplayskip}{3pt}
    K(x)(Z)\le e^{d_\mathcal{X}(x,x')} K(x')(Z) \quad\forall Z\in \mathcal{F_Z}
\label{eq:metric} 
\end{equation}
where $d_\mathcal{X}$ is a distance metric for $\mathcal{X}$, $\mathcal{Z}$ is a set of query outcomes, $\mathcal{F_Z}$ is a $\sigma$-algebra over $\mathcal{Z}$, and $\mathcal{P(Z)}$ is the set of probability measures over $\mathcal{Z}$.

Inspired by geo-indistinguishability \cite{andres2013geo} and image obfuscation \cite{8784836}, which extend metric privacy to location data and images, we design a formal privacy notion for speech data.
As shown in Equation \eqref{eq:metric}, the components of metric privacy are the representation of secrets and the distance between two secrets.
In our case, the secrets are the speakers' voiceprints in a speech database.
We discuss the representation of the voiceprint and the distance metric below.

\textbf{Representation of the voiceprint.} 
The representation of voiceprint is based on models or templates. Studies on voiceprint/speaker recognition algorithms represent voiceprints as features of each vocal cavity, which can fully express the differences of voices. The Gaussian mixture model (GMM) super-vector \cite{svmgmm}, the joint factor analysis (JFA) \cite{jfa}, and the i-vector \cite{ivector} are the main GMM-based methods. When a DNN is used for voiceprint/speaker recognition, the GMM is replaced by the output of the DNN (d-vector) \cite{dvector}. The DNN-based embedding (x-vector) \cite{Snyder2018x} is a state-of-the-art technology. 
The x-vector-based speaker verification approach produces utterance-level embeddings even for variable-length speech segments. Thus, we choose the x-vector as the representation of voiceprint.
We use the same setting used by Snyder et al. \cite{Snyder2018x} and present more technical details about the x-vector in Subsection \ref{ssec:sr}. 

\textbf{Definition of the distance.} 
Similar to $l_1$ norm used in differential privacy that measures the similarity between two neighboring databases, we also need a distance metric  over voiceprints.
The distance between x-vectors should be a metric (i.e., it should satisfy triangle inequality) required by metric privacy. 
Cosine distance could be a candidate since it is widely used in measuring the distance between x-vectors \cite{novoselov2018deep}, but  it is not a well-defined distance metric, as it does not satisfy the triangle inequality property.
We propose to use \textit{angular distance} , which satisfies the inequality property and preserves the similarity between voiceprints.

In the following, we propose a voiceprint privacy definition for a single utterance and for a speech database  in Definition 1  and Definitions 2, respectively.

\textbf{Definition 1 (Voice-Indistinguishability, i.e., Voice-Ind)} A mechanism ${{K}}$ satisfies $\epsilon$-voice-indistinguishability if for any output $\tilde{x} $ and any two possible voiceprints $x, x' \in \mathcal{X}$:
\begin{equation}
\setlength{\abovedisplayskip}{3pt}
\setlength{\belowdisplayskip}{3pt}
\begin{split}
&\frac{\Pr(\tilde{x}|x)}{\Pr(\tilde{x}|x')}\le e^{\epsilon d_\mathcal{X}(x, x')} \label{eq:vi}\\
d_\mathcal{X} = &\frac{\arccos(\cos \,similarity < x,x'>)}{\pi}
\end{split}
\end{equation}
where $\mathcal{X}$ is a set of possible voiceprints, $d_\mathcal{X}$ is the angular distance metric, cos $similarity$ is a measure of similarity between two vectors of an inner product space that measures the cosine of the angle between them.

\textbf{Definition 2 (Speech data release under Voice-Ind)} 
For every two neighboring x-vector databases $\mathcal{D},\mathcal{D'}$, only differing in the $i$th x-vector, which are $x$, and $x'$, a mechanism ${K}$ satisfies $\epsilon$-voice-indistinguishability if for all possible perturbed x-vector databases $\tilde{\mathcal{D}}$
\vspace{-11pt}

\begin{equation}
\vspace{-5pt}
\setlength{\abovedisplayskip}{5pt}
\setlength{\belowdisplayskip}{5pt}
\frac{\Pr(\tilde{\mathcal{D}}|\mathcal{D})}{\Pr(\tilde{\mathcal{D}}|\mathcal{D'})}\le e^{\epsilon d(\mathcal{D}, \mathcal{D'})} \label{eq:vi-db}
\end{equation}
where $d(\mathcal{D}, \mathcal{D'}) = d_\mathcal{X}(x, x')$.
It can be easily found that when the database $\mathcal{D}$ approaches $\mathcal{X}$, the above mentioned definitions are the same.

Voice-Indistinguishability guarantees that given the output x-vector $\tilde{x}$, an attacker hardly distinguishes whether the original x-vector is $x$ or $x'$ bounded by $\epsilon d_\mathcal{X}$.
In other words, a lower $\epsilon d_\mathcal{X}$ indicates higher indistinguishability, hence a higher level of privacy. 
The privacy budget value $\epsilon$ globally influences the degree of guaranteed privacy.

\textbf{Mechanism.} According to the definition, we provide a probable mechanism. Given the input x-vector $x_{0}$, the mechanism ${K}$ perturbs $x_{0}$ by randomly selecting an x-vector $\tilde{x}$ in the dataset $\mathcal{D}$ according to certain probability distributions, thus providing plausible deniability for $x_{0}$. 

\textbf{Theorem 1.} A mechanism ${K}$ that randomly transforms $x_0$ to $\tilde{x}$ where $x_0, \tilde{x}\in \mathcal{D}$ according to the following equation, satisfies voice-indistinguishability
\begin{equation}
\notag
\setlength{\abovedisplayskip}{3pt}
\setlength{\belowdisplayskip}{3pt}
\begin{aligned}
\Pr(\tilde{x}|x_0) \propto e^{-\epsilon d_\mathcal{X}(x_0, \tilde{x})} 
\end{aligned}
\vspace{-10pt}
\end{equation}

\begin{figure}[t]
\begin{center}
\includegraphics[width=0.46\textwidth]{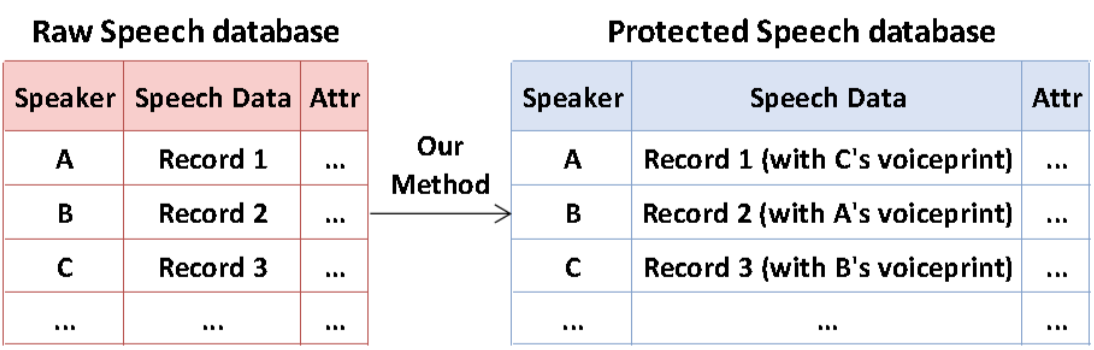}
\vspace{-8pt}
  \caption{Example of a privacy-preserving speech data release.}
  \label{baa}
  \vspace{-20pt}
 \end{center}
\end{figure}

\subsection{Privacy Analysis}

\textbf{Privacy guarantee of the released private speech database}.
Figure \ref{baa} shows an example of the speech database before and after transformation using the proposed mechanism. Voice-indistinguishability guarantees that an attacker can hardly distinguish whether the original voiceprint is from A, B, or C. 

\textbf{Privacy guarantee of voice-indistinguishability.}
We further explain the privacy guarantee provided by voice-indistinguishability by comparing the prior and posterior distributions of information obtained by an adversary.
We prove that the prior and posterior distributions are bounded by $\epsilon d_\mathcal{X}$. In other words, voice-indistinguishability does not impose that an adversary gains no information but limits the increase of information that an adversary can obtain.

Let $\Pr(x)$ and $\Pr(x\mid \tilde{x})$ be the prior and posterior distributions of information obtained by an adversary, respectively, then for two indistinguishable x-vectors $x, x'$:
\begin{equation}
\setlength{\abovedisplayskip}{3pt}
\setlength{\belowdisplayskip}{3pt}
\begin{split}
\lg{\frac{\Pr(\tilde{x}|x)}{\Pr(\tilde{x}|x')}} = \lg{\frac{\Pr(x\mid \tilde{x})}{\Pr(x'\mid \tilde{x})}}-\lg{\frac{\Pr(x)}{\Pr(x')}}\le\,\,\epsilon d_\mathcal{X}(x, x')
\end{split}
\notag
\end{equation}
\vspace{-15pt}

\section{Private Speech Data Release Framework}

We adopt a framework that is similar to speaker anonymization using the x-vector proposed by Fang et al. \cite{fang_speaker_2019}, which is based on feature-level perturbation. However, this framework is time-consuming for a large speech database because the online perturbation time with $n$ records is $O(n^2)$. To improve the time complexity, we propose a Model-level framework. 

\begin{figure}[t]
  \centering
  \includegraphics[width=0.4\textwidth]{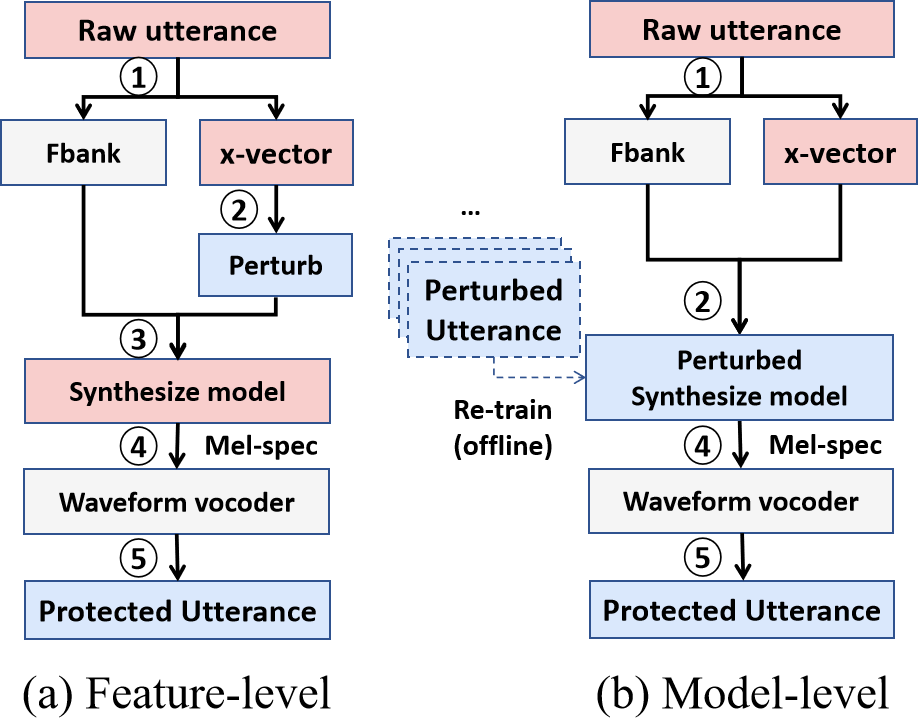}
    \vspace{-10pt}
  \caption{The proposed frameworks.}
  \label{fig:fang}
  \vspace{-10pt}
\end{figure}

Figure \ref{fig:fang} shows these two frameworks. Both frameworks use two modules to generate the speech data: An end-to-end acoustic model that generates a Mel-spectrogram (Mel-spec) (used as a standard input feature in speech synthesis) given the two input features; filter-bank (Fbank) (a commonly used feature for speech recognition) and an x-vector.

\begin{algorithm}[h]
\footnotesize
\caption{Feature-level framework} 
\hspace*{0.02in} {\bf Input:} 
Raw utterance $u_0$,
\hspace*{0.36in}       Utterance database $\mathbb{D}$,\\
\hspace*{0.36in}       Synthesize model $Syn$\\
\hspace*{0.02in} {\bf Output:} 
Protected utterance $\tilde{u_0}$
\begin{algorithmic}[1]
\State Extract Fbank $fbank_0$ and x-vector $x_0$ from $u_0$ \hfill{\textit{Fig.3(a)\textcircled{\scriptsize 1}}}
\For {each utterance $u_i \in \mathbb{D}$}
    \State Extract x-vector $x_i$ from $u_i$
    \State Compute $\Pr(x_i|x_0)$
\EndFor
\State Select x-vector $\tilde{x}$ randomly according to $\Pr(\tilde{x}|x_0)$ \hfill{\textit{Fig.3(a)\textcircled{\scriptsize 2}}}
\State $Mel$-$spec$ ${\gets}$ $Syn$($fbank_0$, $\tilde{x}$) \hfill{\textit{Fig.3(a)\textcircled{\scriptsize 3}}}
\State $\tilde{u_0}$ ${\gets}$ Waveform vocoder($Mel$-$spec$) \hfill{\textit{Fig.3(a)\textcircled{\scriptsize 4}}}
\State \Return $\tilde{u_0}$ \hfill{\textit{Fig.3(a)\textcircled{\scriptsize 5}}}
\end{algorithmic}
\end{algorithm}
\vspace{-15pt}
\begin{algorithm}[h]
\footnotesize
\caption{Model-level framework} 
\hspace*{0.02in} {\bf Input:} 
Raw utterance $u_0$,
\hspace*{0.36in}       Utterance database $\mathbb{D}$,\\
\hspace*{0.36in}       Perturbed synthesize model $\tilde{Syn}$\\
\hspace*{0.02in} {\bf Output:} 
Protected utterance $\tilde{u_0}$
\begin{algorithmic}[1]
\State Extract Fbank $fbank_0$ and x-vector $x_0$ from $u_0$ \hfill{\textit{Fig.3(b)\textcircled{\scriptsize 1}}}
\State $Mel$-$spec$ ${\gets}$ $\tilde{Syn}$($fbank_0$, $x_0$)
\hfill{\textit{Fig.3(b)\textcircled{\scriptsize 2}}}
\State $\tilde{u_0}$ ${\gets}$ Waveform vocoder($Mel$-$spec$) \hfill{\textit{Fig.3(b)\textcircled{\scriptsize 3}}}
\State \Return $\tilde{u_0}$ \hfill{\textit{Fig.3(b)\textcircled{\scriptsize 4}}}
\end{algorithmic}
\end{algorithm}

The only difference between the Feature-level framework and the Model-level framework is seen when we perturb the x-vector. 
As shown in Algorithm 1, Figures \ref{fig:fang}(a) directly add the perturbation to the x-vector before voice synthesis, which represents a Feature-level framework. 
As shown in Algorithm 2, Figures \ref{fig:fang}(b) perturb all utterances, then re-train the synthesis model, so that it learns the perturbation rules, which represents a Model-level framework.

\section{Experiments}

\subsection{Experimental Setting}

We verify the accuracy of voice-indistinguishability in this section. 
Because $\epsilon$ represents our privacy budget, modified speech data with a larger $\epsilon$ has a weaker capacity for mitigating speaker verification attacks but better utility.
Both objective and subjective evaluations are carried out. 

The proposed frameworks are built using the End-to-End speech synthesis toolkit \cite{watanabe2018espnet} on default settings\footnote{https://github.com/espnet/espnet/tree/master/egs/librispeech/tts1}. 
We use the evaluation set of the Librispeech dataset \cite{librispeech} (test-clean) to simulate the proposed frameworks. This dataset consists of 5 hours, 24 min of data, and a total of 40 speakers consisting of 20 males and 20 females. 

For objective evaluation, we test the difference between the original voices and the modified voices using the Mean Squared Error (MSE), and whether they are recognized as the same person using speaker verification as discussed in Subsection \ref{ssec:sr}. In Subsection \ref{ssec:asr}, we also test whether a speech recognition system, which is a utility of speech data, can still recognize the modified speech.
For subjective evaluation, we invite 15 listeners to evaluate the differences between the original and modified speaker voices, and the naturalness of the modified voices. This evaluation is described in Subsection \ref{sec:subj}.
Finally, we compare the two frameworks in Subsection \ref{ssec:com}.

\subsection{Objective Evaluation} 

\begin{table}[t]
\footnotesize
\centering
\begin{tabular}{c|ccc}
Layers & Layer context  & \#context & \#units \\
 \hline
 time-delay 1     & $[t-2,\ t+2]$      & 5             & 512 \\
  time-delay 2     & $\{t-2,\ t,\ t+2\}$& 9             & 512 \\
  time-delay 3     & $\{t-3,\ t,\ t+3\}$& 15            & 512 \\
  time-delay 4     & $\{t\}$            & 15            & 512 \\
  time-delay 5     & $\{t\}$            & 15            & 1500 \\
 \hline
 statistics pooling & $[0,T)$    & $T$     & 3000 \\
 \hline
 bottleneck 1& $\{0\}$               & $T$     & 512 \\
 bottleneck 2& $\{0\}$               & $T$     & 512 \\
 softmax     & $\{0\}$               & $T$  & $L$  \\
\end{tabular}
\caption{The x-vector TDNN. $T$~is the number of frames in a given utterance. $L$~is the number of speakers.}
\vspace{-10pt}
\label{tab:x-vectors}
\end{table}

\subsubsection{Speaker Verification System and Evaluation}
\label{ssec:sr}

We use the open-source x-vector system\footnote{http://kaldi-asr.org/models/m8} pre-trained on the augmented VoxCeleb-1 \cite{voxceleb1} and VoxCeleb-2 datasets \cite{voxceleb2}. The PLDA model is used to evaluate the similarities between the original voices and their modifications. The PLDA model is trained using the speakers in the wild (SITW) dataset \cite{wild}. 

Table \ref{tab:x-vectors} shows the architecture of the x-vector extractor. It consists of the context-aggregating time-delay neural network (TDNN) \cite{tdnnjhu} layers operating at frame level (with the final context window of $\pm$7 frames), a statistics pooling layer which computes the mean and standard deviation of all the frames, effectively changing the variable-length sequence of frame-level activations into a fixed-length vector, and an utterance-level part consisting of two fully connected bottleneck layers which extract more sophisticated features and compress the information into a lower-dimensional space, and an additional softmax output layer.

We evaluate the effect of the size of the speech database and the privacy budget $\epsilon$ on the changes between the original voices and the modified voices.
The Mean Squared Error (MSE) and the accuracy of speaker verification through PLDA (ACC) are used. 
From the results shown in Figure \ref{Vary2}, we can conclude that the capacity for mitigating speaker verification attacks is higher with a larger speech database size and lower $\epsilon$, which means higher level of guaranteed privacy.

\begin{figure}[t]
\centering
  \includegraphics[width=0.44\textwidth]{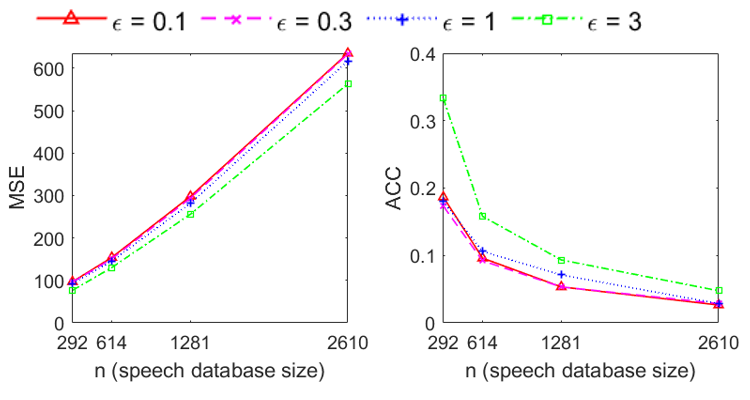}
  \vspace{-10pt}
  \label*{}
\end{figure}
  \vspace{-13pt}
\begin{figure}[t]
\centering
  \includegraphics[width=0.44\textwidth]{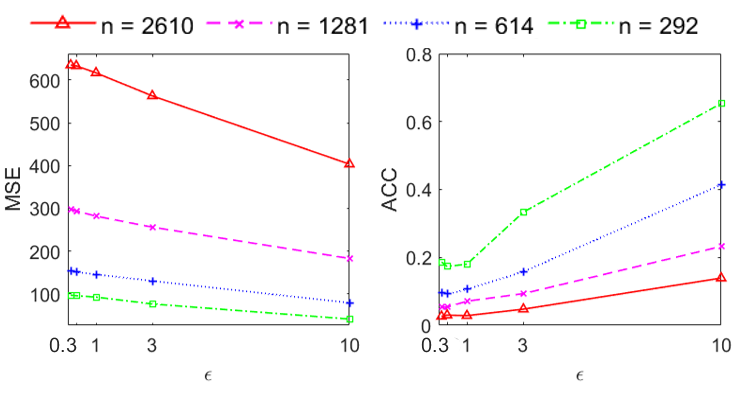}
  \vspace{-10pt}
  \caption{Vary $n$ (speech database size) and $\epsilon$ (privacy budget) }
  \label{Vary2}
  \vspace{-10pt}
\end{figure}

\subsubsection{Speech Recognition System and Evaluation}
\label{ssec:asr}
We use the Mozzilla DeepSpeech framework\footnote{https://github.com/mozilla/DeepSpeech} to construct the speech recognition system. 
We train a three-layer bidirectional long short term memory (BLSTM)-based neural network using the training set of Librispeech (train-clean-100)\footnote{http://www.openslr.org/11/}. 
The model is trained using connectionist-temporal-classification (CTC \cite{ctc}) objective and character-level labels. 
The words in the labels are converted into 26 characters and 3 special tokens (``''' mark, space and blank). This network uses the 161 dim spectrogram features extracted from waveform files. 
Each layer uses 100 gated cells and the ReLU activation function.
The Batch-normalization and dropout techniques (drop-rate=0.48) are used between layers to ensure training stability. 
The output layer has 29-dimensional outputs representing the 29 characters in the label.

\begin{figure}[h!tb]
  \centering\includegraphics[width=0.21\textwidth]{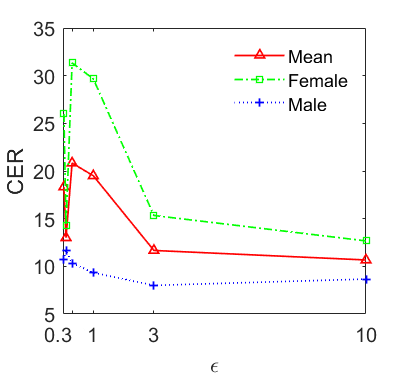}
  \vspace{-10pt}
  \caption{Character Error Rate (CER)}
  \label{Cer}
\end{figure}

For testing, the character level accuracy is calculated. We use the character error rate (CER) to evaluate the performance of speech recognition for protected speech data after perturbing the voiceprint. Figure \ref{Cer} shows a lower CER with a larger $\epsilon$ that is consistent with our definition that a larger $\epsilon$ results in better utility. Note that a small $\epsilon$ may have the same performances because our mechanism is based on randomness. 
\vspace{-5pt}
\vspace{-25pt}

\subsection{Subjective Evaluation}
\label{sec:subj}

\begin{figure}[h!tb]
\centering
  \includegraphics[width=0.43\textwidth]{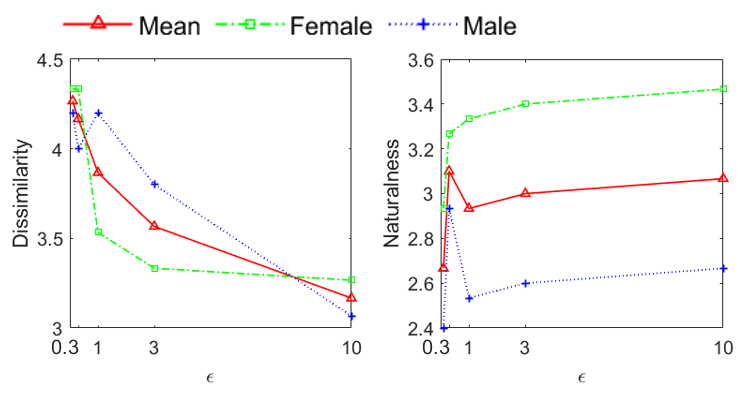}
  \vspace{-10pt}
  \caption{Dissimilarity and Naturalness}
  \label{Mos}
  \vspace{-10pt}
\end{figure}

We invite 15 listeners to measure the speaker's differences between the original voices and the modified voices and the naturalness of sounds that closely resemble the human voice. 
The MOS (Mean Opinion Score) is used to evaluate the generated sound. 
In this study, dissimilarity means the degree of privacy that we protect while naturalness means the utility of the modified speech data. 
As shown in Figure \ref{Mos}, the result of MOS is in accordance with the objective evaluation.

\subsection{Comparison of the Proposed Frameworks}

Based on the same mechanism, the proposed frameworks, the Feature-level framework, and the Model-based framework share similar performances as discussed above. However, without perturbation during testing, the time complexity of the Model-level framework shows a reduction of $O(n^2)$. 

As shown in Table \ref{time}, the Feature-level framework using online perturbation time is time-consuming when $n$ is large. This proves that the Model-level framework outperforms the Feature-level framework in time complexity.

\begin{table}[t]
\small
\centering
\begin{tabular}{c|c|c|c}
Speech database size & $n$ = 100  & $n$ = 1000 & $n$ = 10000\\
 \hline
Perturbation time     & 0.4802s      & 29.0959s             & 2710.9289s 
\end{tabular}
\caption{The online perturbation time.}
\vspace{-15pt}
\label{time}
\end{table}
\label{ssec:com}

\section{Related works}

Speaker de-identification \cite{justin2015speaker,qian2018hidebehind,srivastava2019evaluating,fang_speaker_2019, Srivastava2019} seeks to change the identity information without affecting the textual content.
The existing speaker de-identification algorithms include:

\textbf{(1) Voice-level Protection}.
Voice-level Algorithms are based on voice conversion. 
They construct a mapping from the source voice to the target voice. 
VoiceMask as described by Qian et al., \cite{qian2018hidebehind} does such mapping based on vocal tract length normalization (VTLN), which is a well-studied voice conversion technology that uses frequency warping. 
Srivastava et al. \cite{srivastava2019evaluating} propose another VTLN-based approach that warps the frequency axis in different directions over time. 
It can be concluded that both of them provide no intuitive definition of voiceprint.

\textbf{(2) Feature-level Protection}.
Feature-level algorithms are based on voice synthesis. Justin et al. \cite{justin2015speaker} propose an approach that transforms all the speakers' voices to that of a specific speaker.
Similarly, speaker anonymization using the x-vector and neural waveform models is discussed by Fang et al. \cite{fang_speaker_2019}. The main assumption is that speech waveform can be anatomized by altering the features that encode the speaker's identity.
These two methods indeed protect the voiceprint of voices but do not point out the degree of voiceprint protection that they provide.

\textbf{(3) Model-level Protection}.
Srivastava et al. \cite{Srivastava2019} point out that the content feature obtained from a standard automatic speech recognition (ASR) architecture still carries abundant information regarding speaker identity. 
They use adversarial training to learn the content feature while hiding speaker identity.
However, their approach has a limited effect on the accuracy of speaker validation.

\section{Conclusion and Future Work}

This study proposes a privacy definition for voiceprint, voice-indistinguishability, and present a speaker de-identification solution using two frameworks.
Consequently, our solution achieves a balanced trade-off between privacy and utility. 
Both objective and subjective evaluations prove that our method can mitigate speaker verification attacks and achieve an improved performance of speech recognition.
In future studies, we shall propose mechanisms with better utility and broadly evaluate the applicability of the frameworks.

\vspace{-5pt}
\section{acknowledgments}

This work is partially supported by JSPS KAKENHI Grant No. 17H06099, 18H04093, 19K20269, 19K24376, NICT tenure-track startup fund, and Microsoft Research Asia (CORE16).
\vspace{-5pt}

\bibliographystyle{IEEEbib}
\bibliography{CR.bib}

\end{document}